\documentclass[12pt]{article}
\usepackage{amssymb,amsmath,latexsym,theorem}
\def \slas{\kern -7.2pt /}
\def \sla{\kern -5.4pt /}
\def \Cslas{\kern -6.8pt /}
\def \Dslas{\kern -7.4pt /}
\def \ii{\mathrm{i}}
\def \d{\mbox{d}}

\def \pd{\partial}

\def \tl#1{\overset{\kern 0pt\circ}{#1}}
\def \TL#1{\overset{\kern -3pt \circ}{#1}}

\def \TLL#1{\overset{\kern -7pt \circ}{#1}}
\def \TLLL#1{\overset{\kern -30pt \circ}{#1}}

\newcommand{\bq}{\begin{equation}}
\newcommand{\eq}{\end{equation}}
\newcommand{\bea}{\begin{eqnarray}}
\newcommand{\eea}{\end{eqnarray}}

\newcommand\ka{\kappa_1}
\newcommand\kb{\kappa_2}

\newcommand\xx{\tilde{x}}

\normalbaselines
\begin{document}
\begin{center}
\vspace*{1.0cm}

{\LARGE{\bf {Twist Decomposition of Nonlocal\\[1mm] Light-Ray Operators
and Harmonic\\[2mm] Tensor Functions\footnote{
Contribution to Proceedings of International
Symposium "Quantum Theory and
Symmetries", Goslar, 18. -- 22. July 1999}
}}}

\vskip 1cm

{\large{\bf B. Geyer} ~and~ {\bf M. Lazar}} 

\vskip 0.5 cm

Naturwissenschaftlich-Theoretisches Zentrum, \\
Universit{\"a}t Leipzig, Augustusplatz 10, D--04109 Leipzig,
Germany 
\end{center}
%\vspace{.5cm}
%\begin{abstract}
%text of Abstract
%\end{abstract}
%\vspace{.5 cm}

\section{Introduction}

Light--cone(LC) dominated hard scattering processes, like deep
inelastic scattering
and deeply virtual Compton scattering, and various hadronic
wave functions, e.g.,~vector meson amplitudes, appearing in (semi) exclusive
processes are most effectively described  by the help of the {\em nonlocal}
light--cone expansion \cite{AZ}.Thereby, the {\em same} nonlocal LC--operator,
and its anomalous dimension, is related to {\em different} phenomenological
distribution amplitudes, and their $Q^2$--evolution kernels \cite{MRGDH}.
Growing experimental precision requires the consideration of higher twist
contributions. Therefore, it is necessary to decompose the nonlocal
LC--operators according to their twist content.

Unfortunately, `geometric' twist  $(\tau) =$ dimension $(d)-$ spin $(j)$,
introduced for the {\em local} LC--operators \cite{GT} cannot be
extended directly to nonlocal LC--operators. Therefore, motivated
by LC--quantization %\cite{KS}
and by kinematic phenomenology the notion of `dynamic' twist $(t)$ was
introduced counting powers $Q^{2-t}$ of the momentum transfer \cite{JJ}.
However, this notion is defined only for
{\em matrix elements} of operators, is {\em not} Lorentz invariant and
its relation to `geometric' twist is complicated, cf.~\cite{BBKT}.
%This, at least, may led to phenomenological inaccuracies.

Here, we carry on a systematic procedure to uniquely decompose nonlocal
LC--operators into harmonic operators of well defined geometric twist,
cf.~Ref.~\cite{GLR}. This will be demonstrated for tensor operators
of low rank, namely (pseudo)scalars, (axial) vectors and (skew)symmetric tensors
Thereby, various harmonic tensor operators (in $D= 2h$ space-time dimensions)
are introduced being related to specific infinite series of
symmetry classes which are defined by corresponding Young tableaus.
In the scalar case these LC--operators are series of  the well-known harmonic
polynomials corresponding to (symmetric) tensor representations of the
orthogonal group $SO(D)$, cf.~\cite{BT}. %Bargmann & Todorov.
Symmetric tensor operators of rank 2 are considered as an example.

\section{General procedure}

\baselineskip=11pt

Let us generically denote by $O_\Gamma(\ka \xx, \kb \xx)$
%\equiv \phi^a(\ka \xx)U^{ab}(\ka \xx, \kb \xx)\phi^b(\kb \xx)$
with $\xx^2 = 0$ any of the following bilocal light-ray operators,
 either the quark operators
$:\overline\psi(\ka \xx) \Gamma U(\ka \xx, \kb \xx) \psi(\kb \xx):$
with $\Gamma=\{1,\gamma_\alpha,\sigma_{\alpha\beta},
\gamma_5\gamma_\alpha,\gamma_5\}$ or the gluon operators
$:{F_\mu}^\rho(\ka \xx) U(\ka \xx, \kb \xx) F_{\nu\rho}(\kb \xx):$
and
${:{F_\mu}^\rho(\ka \xx) U(\ka \xx, \kb \xx)
{\widetilde F}_{\nu\rho}(\kb \xx):}$, where
$\psi(x), F_{\mu\nu}(x)$ and ${\widetilde F}_{\mu\nu}(x)$ are
the quark field, the gluon field strength and its dual, respectively,
and the path ordered phase factor is given by
$U(\ka x, \kb x)={\cal P} \exp\{-\ii g
\int^{\ka}_{\kb} d\kappa \,x^\mu A_\mu (\kappa x)\}$
with the gauge potential $A_\mu(x)$.
The general procedure of decomposing these operators
into operators of definite twist  consists of the following steps:\\
(1) {\em Expansion} of the nonlocal operators for {\em arbitrary}
values of $x$
into a Taylor series of {\em local} tensor operators
having definite rank $n$ and canonical dimension $d$:
\begin{eqnarray}
\label{GLCEOP}
O_\Gamma(\ka x, \kb x)=\hbox{$\sum_{n=0}^{\infty}$}
(n!)^{-1}{x}^{\mu_1}\ldots{x}^{\mu_n}
O_{\Gamma\,\mu_1\ldots\mu_n}(\ka,\kb);
\end{eqnarray}
the local operators are defined by
(the brackets $(\ldots)$ denote total symmetrization)
\begin{eqnarray}
\label{local}
O_{\Gamma\,\mu_1\ldots\mu_n}(\kappa_1,\kappa_2)
\equiv
\left[\Phi(x)\Gamma
{\sf D}_{(\mu_1}\!\ldots{\sf D}_{\mu_n)}
\Phi(x)\right]\big|_{x=0},
%\nonumber
\end{eqnarray}
with generalized covariant derivatives
%\begin{eqnarray}
%\label{D_kappa}
${\sf D}_\mu(\kappa_1,
\kappa_2)\equiv
\kappa_1 (\overleftarrow{\pd_\mu}-\ii g A_\mu)+
\kappa_2 (\overrightarrow{\pd_\mu}+\ii g A_\mu).
%\nonumber
%\end{eqnarray}
$

\noindent
(2) {\em Decomposition} of the local operators (\ref{local}) into
tensors being {\em irreducible} under the (ortho\-chronous)
Lorentz group, i.e.~having twist $\tau = d-j$. These are
{\em traceless} tensors being classified
according to their {\em symmetry class}. The latter
 is determined by a (normalized) {\em  Young-Operator}
 ${\cal Y}_{[m]}={f_{[m]}}{\cal QP}/{n!}$,
where $[m]= (m_1\geq m_2\geq\ldots\geq m_r)$ with
$\sum^r_{i=1}m_i = n$ defines a Young pattern, and
${\cal P}$ and ${\cal Q}$ denote symmetrizations and
antisymmetrizations w.r.to the horizontal and vertical
permutations of a standard tableau, respectively;
$f_{[m]}$ is the number of standard tableau's to $[m]$.

This decomposition can be done for any dimension $D = 2h$
of the complex orthogonal group $SO(2h, C)$.
Then, the allowed Young patterns are restricted by
$\ell_1+\ell_2 \leq 2h~ (\ell_i$: length of columns of $[m]$).
Since the operators (\ref{local}) are totally symmetric w.r.t.
$\mu_i$'s only the following symmetry types, %(i) -- (iv),
depending
on the additional tensor structure $\Gamma$,
are of relevance (lower spins $j$ are related to trace terms):\\
$
\begin{array}{lllll}
\hspace{.5cm}
\rm{(i)}&[m] = (n)&j = n, n-2, n-4,..., & f_{(n)} &= 1,\\
\hspace{.5cm}
\rm{(ii)}&[m] = (n,1)&j = n, n-1, n-2,..., & f_{(n,1)}&= n,\\
\hspace{.5cm}
\rm{(iii)}&[m] = (n,1,1)&j = n, n-1,n-2,..., & f_{(n,1,1)} &= n(n+1)/2,\\
\hspace{.5cm}
\rm{(iv)}&[m] = (n,2)&j =  n, n-1,n-2,..., & f_{(n,2)} &= (n+1)(n+2)/2.\\
\end{array}
$
\vspace*{0cm}

\noindent
If multiplied by ${x}^{\mu_1}\ldots{x}^{\mu_n}$ according to
Eq.~(\ref{GLCEOP}) these representations may be characterized by
harmonic tensor polynomials of order $n$, cf.~Sect. 3 below.\\
%
%Because the decomposition into irreducible
%representations does not depend on the specific values of
%the $\kappa$--variables we may choose
%$(\kappa_1,\kappa_2)=(0,\kappa)$ and
%${\sf D}_\mu(\kappa_1, \kappa_2)
%Rightarrow \kappa {\stackrel{\rightarrow}{D}}_\mu (A)$;
%afterwards we generalize to arbitrary values of $\kappa_i$.
%
(3) {\em Resummation} of the infinite series (for any $n$)
of irreducible tensor harmonics of equal twist $\tau$ and symmetry type
 ${\cal Y}_{\rm [m]}$ to nonlocal {\em  harmonic operators of definite twist}.
Thereby, the phase factors may be reconstructed on the expense of (two)
additional integrations over parameters multiplying the $\kappa$--variables.
As a result one obtains a decomposition of the original nonlinear operator
into a series of operators with growing twist, each being defined by subtracting
appropriate traces, cf. Sect. 4 below.\\
(4) {\em Projection} onto the light--cone, $x\rightarrow\xx = x + \eta
{(x\eta)}\big(\sqrt{1 - {x^2}/{(x\eta)^2}} - 1\big),~~ \eta^2=1$,
leads to the required twist decomposition:
\begin{eqnarray}
\label{LCdecomp}
O_\Gamma(\kappa_1 \tilde x,\kappa_2\tilde{x})
~=&
\sum_{\tau_i = \tau_{\rm min}}^{\tau_{\rm max}}
O_\Gamma^{\tau_i}(\kappa_1 \tilde x,\kappa_2\tilde{x}).
%\nonumber
\end{eqnarray}
This sum terminates since most of the traces are proportional to
$x^2$ and, therefore, vanish on the light--cone. \\
Let us remark that step (3) and (4) may be interchanged without
changing the result. In fact, this corresponds to another formalism
acting on the complex light--cone \cite{BT} being used for the
construction of conformal tensors by Dobrev and Ganchev \cite{DG}.

\section{Harmonic tensor polynomials}

The construction of irreducible tensors may start with
traceless tensors $\tl T_{\Gamma(\mu_1 \ldots \mu_n)}$, with
$\Gamma$ indicating additional tensor indices, which afterwards
have to be subjected to the symmetry requirements determined
by the symmetry classes (i) -- (iv).
Let us start with a generic tensor, not being traceless, whose
symmetrized indices are contracted by $x^{\mu_i}$'s:
 $T_{\Gamma n}(x) =
x^{\mu_1}\ldots x^{\mu_n} T_{\Gamma (\mu_1 \ldots \mu_n)}$.
The conditions
for scalar, (axial) vector resp.~(skew)symmetric tensors
to be traceless read:
\begin{eqnarray}
\hspace{-.5cm}
\square \tl T_{\Gamma~n}(x)&=&0,
\\
\hspace{-.5cm}
\pd^\alpha \tl T_{\alpha~n}(x) = 0
 \quad{\rm resp.}\quad
\pd^\alpha \tl T_{\alpha\beta~n}(x) =\!\!\! &0&\!\!\! =
\pd^\beta \tl T_{\alpha\beta~n}(x)
\quad{\rm and}\quad
g^{\alpha\beta} \tl T_{\alpha\beta~n}(x) = 0.
%\nonumber
\end{eqnarray}
The general solutions of these equations in $D=2h$ dimensions are:\\
(1) {\em Scalar harmonic polynomials} (cf.~\cite{VK}):
\begin{eqnarray}
\label{T_harm4}
\tl T_n(x) =
\sum_{k=0}^{[\frac{n}{2}]}
\hbox{\large$\frac{(h+n-k-2)!}{k!(h+n-2)!}$}
\left(\hbox{\large$\frac{-x^2}{4}$}\right)^{\!\!k}
\square^{k}T_n(x)
\equiv
H^{(2h)}_n\!\left(x^2|\square\right)T_n(x)
\end{eqnarray}
(2) {{\em Vector harmonic polynomials}:
\begin{eqnarray}
\label{T_vecharm}
\hspace{-.5cm}
\tl T_{\alpha n}(x)
\!\!\!\!&=&\!\!\!\!
\Big\{\!\delta_{\alpha}^{\beta}
\!-\!\hbox{\large$\frac{1}{(h+n-1)(2h+n-3)}$}
\Big(\!(\!h\!-\!2\!+\!x\pd)x_\alpha\pd^\beta
\!-\!\hbox{\large$\frac{1}{2}$}x^2
\pd_\alpha\pd^\beta\Big)\!\Big\}\!
H^{(2h)}_n\!\left(x^2|\square\right)\! T_{\beta n}(x)
\end{eqnarray}
(3) {{\em Skew tensor harmonic polynomials}:
\begin{eqnarray}
\label{T_tenharm}
\hspace{-.5cm}
\lefteqn{
\tl T_{[\alpha\beta]n}(x)
=
\Big\{\!\delta_{[\alpha}^\mu\delta_{\beta]}^\nu
+\hbox{\large$\frac{2}{(h+n-1)(2h+n-4)}$}
\Big(\!(\!h\!-\!2\!+\!x\pd)
x_{[\alpha}\delta_{\beta]}^{[\mu}\pd^{\nu]}
-\hbox{\large$\frac{1}{2}$}x^2
\pd_{[\alpha}\delta_{\beta]}^{[\mu}\pd^{\nu]}\Big)}
\nonumber\\
\hspace{-.5cm}
&\qquad\qquad
-\hbox{\large$\frac{2}{(h+n-1)(2h+n-4)(2h+n-2)}$}
x_{[\alpha}\pd_{\beta]}x^{[\mu}\pd^{\nu]}
\Big\}
H^{(2h)}_n\!\left(x^2|\square\right)
T_{[\mu\nu] n}(x)
\end{eqnarray}
(4) {\em Symmetric tensor harmonic polynomials}:
\begin{eqnarray}
\label{T_symharm}
\hspace{-.5cm}
\lefteqn{
\tl T_{(\alpha\beta)n}(x)
=
\Big\{\delta_{(\alpha}^\mu\delta_{\beta)}^\nu
+a_n g_{\alpha\beta}x^{(\mu}\pd^{\nu)}
-2g_n x_{(\alpha}\delta_{\beta)}^{(\mu}\pd^{\nu)}
-2b_n x_{(\alpha}x^{(\mu}\pd_{\beta)}\pd^{\nu)}
+c_n x^2 \delta_{(\alpha}^{(\mu}\pd_{\beta)}\pd^{\nu)}
}
\nonumber\\
\hspace{-.5cm}
&\qquad\qquad%\quad
-d_n x^2 x_{(\alpha}\pd_{\beta)}\pd^\mu\pd^\nu
+e_n x_\alpha x_\beta \pd^\mu \pd^\nu
+f_n x^2 x^{(\mu}\pd^{\nu)}\pd_\alpha\pd_\beta
-k_n x^2 g_{\alpha\beta} \pd^\mu\pd^\nu
\nonumber\\
\hspace{-.5cm}
&\qquad\quad\;\;\;
+h_n (x^4/4) \pd_\alpha\pd_\beta\pd^\mu\pd^\nu \Big\}
\Big(\delta_\mu^\rho\delta_\nu^\sigma
- (2h)^{-1} g_{\mu\nu}g^{\rho\sigma}\Big)
H^{(2h)}_n\!\left(x^2|\square\right)
T_{(\rho\sigma) n}(x)
\end{eqnarray}
where, up to the common factor
$[(h-1)(h+n)(h+n-1)(2h+n-2)(2h+n-3)]^{-1}$,
 the coefficients are given by\\
$
\begin{array}{lll}
a_n = (h+n-1)(h+n-2)(2h+n-3), &\qquad
b_n = (h+n-2)(2h+n-3),\\
c_n = [h(h+n-1) -n+2](2h+n-3), &\qquad
d_n = [h(h+n-4)-2(n-3)],\\
\end{array}
$
\\%vspace
$
\begin{array}{lll}
e_n = [h(h+n-2) - n+3](h+n-2), &\qquad
f_n = (2h+n-3), \\
g_n = [h(h+n-1)-n+1](h+n-2)(2h+n-3), &\qquad
h_n = (h-3) \\
k_n = [h(h+n-3) -2n+3]+(h+n)(h+n-1)/2.&\\
\end{array}
$\\
%with the convention
%$T_{[\alpha \beta]} = (T_{\alpha\beta} - T_{\beta\alpha})/2$.\\
% {\em harmonic tensor function} may be represented
%as a series expansion into corresponding polynomials.
These coefficients look quite complicated. However, in $D=4$
dimensions they simplify considerably. The extension to tensors of
higher rank is obvious, but cumbersome. Up to now no simple
building principle
has been found, and a general theory, to the best of our knowledge,
does not exist.

\section{Twist decomposition: An example}

Besides the scalar case these harmonic tensor
polynomials are not irreducible with respect to $SO(2h, C)$. Irreducible
polynomials are obtained by (anti)symmetrization according to the
Young patterns (i) -- (iv); this may be achieved by acting on it with
corresponding differential operators, multiplied by the weights $f_{[m]}$.
Afterwards, in order to get the (local) operators on the light--cone,
one takes
the limit $x \rightarrow \xx$. The nonlocal LC--operators are obtained
by resummation according to Eq.~(\ref{GLCEOP}).

For the quark operators that procedure has been already considered in
\cite{GLR}. Here we shall consider the gluon operators being tensors
of rank 2. The relevant Young patterns are given by (i) -- (iv).
Of course, in physical applications they appear sometimes also
multiplied with $x_\alpha$ or/and $x_\beta$; then higher patterns
are irrelevant.

Let us consider totally symmetric tensors
leading to much simpler symmetric harmonic tensor
polynomials then those given by the general expression (\ref{T_symharm}):
\begin{align}
\label{sy_ha_te_po}
T^{\rm (i)}_{(\alpha\beta) n}(x)=
&\,\hbox{\large$\frac{1}{(n+2)(n+1)}$}\pd_\alpha\pd_\beta
\tl T_{n+2}(x)%\nonumber\\
=\,\hbox{\large$\frac{1}{(n+2)(n+1)}$}
\sum_{k=0}%^{\left[\frac{n+2}{2}\right]}
\hbox{\large$\frac{(-1)^k(h+n-k)!}{4^kk!(h+n)!}$}
\Big\{
x^{2k}\pd_\alpha\pd_\beta %T_{n+2}(x)
\nonumber\\
&+
2k x^{2(k-1)}\big(g_{\alpha\beta}+2 x_{(\alpha}\pd_{\beta)}\big)
%\square^{k} T_{n+2}(x)
%\nonumber\\&
+ 4k(k-1)x_\alpha x_\beta x^{2(k-2)}\Big\}\square^{k}
 T_{n+2}(x).
\end{align}
Related to the number of partial derivatives this sum contains
symmetric tensor, vector and scalar polynomials of lower degree.
Of course, if projected onto the light-cone only those terms survive which
do not depend on $x^2$. They are given by
\begin{align}
\label{sy_te_po(z)}
T^{\rm (i)}_{(\alpha\beta) n}(\xx)
=&\hbox{\large$\frac{1}{(n+2)(n+1)(h+n)(h+n-1)}$}
\d_\alpha\d_\beta T_{n+2}(\xx),
\end{align}
with the `interior' derivative
$\d_\mu=[(h-1+x\pd)\pd_\mu -\frac{1}{2}x_\mu\square]\big|_{x = \xx}$
on the light-cone (cf. Ref.~\cite{BT}); because of $\d^2=0$
the conditions of tracelessness on the light-cone,
$%\begin{align}
\d^\alpha T^{\rm (i)}_{(\alpha\beta) n}(\xx)=0=
\d^\beta T^{\rm (i)}_{(\alpha\beta) n}(\xx)$ {and}
$g^{\alpha\beta} T^{\rm (i)}_{(\alpha\beta) n}(\xx)=0,
$ %\end{align}
are trivially fulfilled.

The local gluon operator having twist $\tau = 2h-2$ is obtained
from the original operator $G_{\alpha\beta}(x) =
\frac{1}{(n+2)(n+1)}\pd_\alpha\pd_\beta  G_{n+2}(x)$
by subtracting the traces being of twist $2h$ and $2h+2$. These traces,
corresponding to irreducible representations, are %given by
\begin{align}
\label{yyy}
G^{\rm tw(2h)(i)a}_{(\alpha\beta) n}(\xx)
=& \hbox{\large$ \frac{1}{2(n+2)(n+1)(h+n)}$}g_{\alpha\beta}\square
G_{n+2}(x)\big|_{x=\xx},
\\
G^{\rm tw(2h)(i)b}_{(\alpha\beta) n}(\xx)
=&
\hbox{\large$ \frac{1}{(n+2)(n+1)(h+n)}$}
\Big(x_{(\alpha}\pd_{\beta)}\square
-\hbox{\large$ \frac{1}{2(h+n-2)}$}
x_{\alpha}x_{\beta}\square^2
\Big)G_{n+2}(x)\big|_{x=\xx},
\\
G^{\rm tw(2h+2)(i)a}_{(\alpha\beta) n}(\xx)
=&-\hbox{\large$ \frac{1}{4(n+2)(n+1)(h+n)(h+n-1)}$}
x_{\alpha}x_{\beta}\square^2
G_{n+2}(x)\big|_{x=\xx},
\\
\label{xxx}
G^{\rm tw(2h+2)(i)b}_{(\alpha\beta) n}(\xx)
=&\hbox{\large$ \frac{1}{2(n+2)(n+1)(h+n)(h+n-2)}$}
x_{\alpha}x_{\beta}\square^2 G_{n+2}(x)\big|_{x=\xx}.
\end{align}

For the nonlocal gluon operator
$G_{(\alpha\beta)}(\ka\xx,\kb\xx)=F_{(\alpha}^{\, \rho}(\ka\xx)
U(\ka\xx,\kb\xx) F_{\beta)\rho}(\kb\xx)$
the expressions (\ref{sy_te_po(z)}) may be resummed leading,
if restricted to $D=4$,
to the following one-parameter integral representation
($G(\ka\xx,\kb\xx)\equiv
x^\alpha x^\beta G_{(\alpha\beta)}(\ka\xx,\kb\xx)$):
\begin{align}
\label{Gtw2_gir}
G^{\mathrm{tw2}(\ii)}_{(\alpha\beta)} (\ka\xx,\kb\xx)
=&
\int_{0}^{1}\d\lambda \Big\{(1-\lambda)\pd_\alpha\pd_\beta
-(1-\lambda+\lambda\ln\lambda)
\left(\hbox{\large$\frac{1}{2}$}g_{\alpha\beta}
+x_{(\alpha}\pd_{\beta)}\right)\square
\nonumber\\
&-\hbox{\large$\frac{1}{4}$}\big(2(1-\lambda)+(1+\lambda)\ln\lambda\big)
x_\alpha x_\beta\square^2\Big\}
G(\ka\lambda x,\kb\lambda x)\Big|_{x=\xx}.
%\nonumber
\end{align}
Analogous expressions result for (\ref{yyy}) -- (\ref{xxx}).
This completes the twist decomposition of
$G_{(\alpha\beta)}^{\mathrm{(i)}} (\ka\xx,\kb\xx)$. The decomposition
of $G_{\alpha\beta}(\ka\xx,\kb\xx)$ w.r.to the other symmetry classes is
much more complicated. A complete presentation  will be
given elsewhere.

%\section*{Acknowledgment}
The authors would like to thank D. Robaschik for
many stimulating discussions.

\end{document}